\documentclass[a4paper,amsmath,amssymb,amsfonts,showpacs,showkeys,twocolumn,titlepage,nobalancelastpage,raggedbottom,floatfix]{revtex4}

\usepackage[english]{babel}
\usepackage{ifpdf}
\ifpdf
\usepackage[pdftex]{graphicx}
\usepackage[protrusion=true,expansion=true]{microtype}  
\usepackage[pdftex,colorlinks,linkcolor=black,citecolor=black,urlcolor=black,hyperfootnotes=false]{hyperref}
\else
\usepackage[dvips]{graphicx,color}
\usepackage{microtype}
\usepackage[dvips,colorlinks,linkcolor=black,citecolor=black,urlcolor=black,hyperfootnotes=false]{hyperref}
\fi

\newcommand{\integral}[4]{
\int_{#3}^{#4}\mathrm{d}#2 \, #1 }

\newcommand{\sinb}[1]{ \mathrm{sin}(#1) }

\newcommand{\cosb}[1]{ \mathrm{cos}(#1) }

\newcommand{\degree}[0]{^{\circ}}

\begin{document}
\selectlanguage{english}
\title{A model for the orientational ordering of the plant microtubule cortical array}
\author{Rhoda J. Hawkins}
\altaffiliation[Current address: ]{UMR 7600, Universit\'e Pierre
et Marie Curie/CNRS, 4 Place Jussieu, 75255 Paris Cedex 05 France}
\email{rhoda.hawkins@physics.org}
\thanks{The first two authors contributed equally to this work.}
\author{Simon H. Tindemans}
\thanks{The first two authors contributed equally to this work.}
\author{Bela M. Mulder}
\affiliation{FOM Institute for Atomic and Molecular Physics,
Science Park 113, 1098 XG, Amsterdam, The Netherlands}
\begin{abstract}
The plant microtubule cortical array is a striking feature of all growing plant cells. It consists of a more or less
homogeneously distributed array of highly aligned microtubules connected to the inner side of the plasma membrane
and oriented transversely to the cell growth axis. Here we formulate a continuum model to describe the origin of
orientational order in such confined arrays of dynamical microtubules. The model is based on recent experimental
observations that show that a growing cortical microtubule can interact through angle dependent collisions with
pre-existing microtubules that can lead either to co-alignment of the growth, retraction through catastrophe
induction or crossing over the encountered microtubule. We identify a single control parameter, which is fully
determined by the nucleation rate and intrinsic dynamics of individual microtubules. We solve the model analytically
in the stationary isotropic phase, discuss the limits of stability of this isotropic phase, and explicitly solve for
the ordered stationary states in a simplified version of the model.
\end{abstract}



\maketitle


\section{Introduction}

\label{sec_intro} 
\begin{figure}[hbt]
  \begin{centering}
  \includegraphics{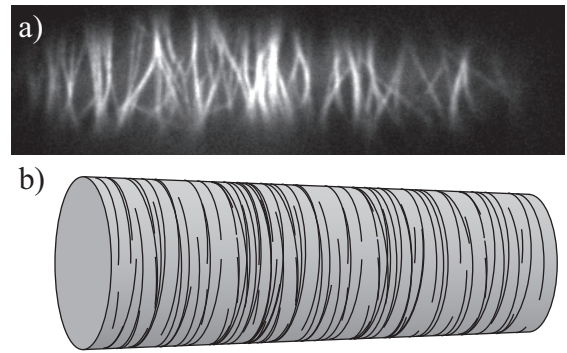}
   \caption{\label{fig_Jan_CA}(a) Fluorescently labelled microtubules in a Tobacco BY-2 cell expressing GFP:TUA6.
   Image courtesy of Jelmer Lindeboom, Wageningen University. (b) Schematic representation of a plant cell cortical array. }
 \end{centering}
\end{figure}

Most plant cells grow by uniaxial expansion. Establishing and maintaining this characteristic
anisotropic growth mode requires regulatory mechanisms that are robust, and, in addition, sensitive to the cell
geometry. A major role in this process is played by microtubules, highly dynamic filamentous protein aggregates that
form one of the components of the cytoskeleton of all eukaryotic organisms (see Ref.\ \cite{Alberts}, chapter 16).
In growing plant cells microtubules are confined to a thin layer of cytoplasm just inside the cell plasma membrane.
Here they form the so-called cortical array, an ordered structure formed by highly aligned (bundles of) microtubules
oriented transversely to the growth direction \cite{EhrhardtShaw06}. This structure is unique to plant cells and
there is evidence that it controls the direction of cell expansion by guiding the mobile transmembrane protein
complexes that deposit long cellulose microfibrils in the plant cell wall \cite{Paradez2006}. These cellulose
microfibrils are the main structural elements of the cell wall, which, for mechanical reasons, are also
transversely oriented to the cell axis in growing cells \cite{cosgrove2005}. An \emph{in vivo} image of the array
and a schematic are shown in Figure~\ref{fig_Jan_CA}.
\emph{In vivo} imaging of microtubules labelled with fluorescent
proteins in plant cells by several groups has shown how the
cortical array is established both following cell division and
after microtubule depolymerizing drug (oryzalin) treatment
\cite{Wasteneys1989,Kum+01,Shaw+03,Vos+03,Paradez2006,EhrhardtShaw06}.
In these studies microtubules are seen to nucleate at the cortex
and then develop from an initially disorganized state into the
transverse ordered array over a time period on the order of one
hour. The nature of the self-organization process by which the
specific spatial and orientational patterning of this cytoskeletal
structure is achieved is as yet only partially understood and
forms the subject of this work.

An important aspect of the problem is the nature of localization
of the microtubules to the cortical region. Fluorescence recovery
after photo-bleaching (FRAP) experiments by Shaw et al.\
\cite{Shaw+03} showed that the microtubules are fixed in space, so
any apparent mobility of microtubules is due to `treadmilling',
the process of simultaneous polymerization at one end and
depolymerization at the other end. So, cortical microtubules do
not translate or rotate as a whole. The same authors also did not
detect detachment or (re)attachment of microtubules to the cell
cortex, apart from some growing ends of single microtubules moving
out of focus and found no evidence for motors working in the
cortical array. These experiments indicate that the microtubules
in the cortical array are fixed to the inside of the cell
membrane. Electron microscopy has also shown cross-bridges between
cortical microtubules and the membrane \cite{Hardham1978}. It is
therefore widely assumed that there are linker proteins that
anchor the microtubules through the plasma membrane to the rigid
cell wall, although their molecular identity is under debate
\cite{Dhonukshe2003,Gardiner2001,HashKato06,Hamada2007,Kirik2007}.
Since the cortical microtubules are effectively confined to a 2D
surface, they can interact through `collisions' that occur when
the polymerizing tip of a growing microtubule encounters a
pre-existing microtubule. The resulting dynamical interaction events
were first characterized by Dixit and Cyr \cite{DixCyr04} in
tobacco Bright Yellow-2 (BY-2) cells. They observed three
different possible outcomes: (i) \emph{zippering}: a growing
microtubule bending towards the direction of the microtubule
encountered, which occurs only when the angle of incidence is
relatively small ($\lesssim 40^{\circ}$) (ii) \emph{induced
catastrophe}: an initially growing microtubule switching to a
shrinking state and retracting after the collision, an effect
predominant at larger angles of incidence and (iii)
\emph{cross-over}: a growing microtubule `slipping over' the one
encountered and continuing to grow in its original direction.

There are clearly many coupled mechanisms at work in this complex
biological system contributing to the assembly and maintenance of
this microtubule cortical array structure. We are interested in
understanding what are the main contributing factors and how their
interplay leads to the observed orientational ordering. With this aim
we develop a coarse-grained model, incorporating all the
effects discussed above. Our emphasis on the plant-specific
biological mechanism of the ordering in the cortical array
distinguishes our approach from earlier work.

Over the years, various models for self-organization of
cytoskeletal filaments (and polar rods in general) have been
proposed \cite{geigant98,kruse05,aranson06,ruehle08}, and the
model by Zumdieck et al. \cite{Zum+05} was applied to the plant
cortex. However, in each of these models the filaments are assumed
to have rotational and, in most cases, translational degrees of
freedom. This is inconsistent with the fact that the plant
cortical microtubules are stably anchored. Inspired by the
experimental results of Dixit and Cyr \cite{DixCyr04}, Baulin et
al.\ \cite{Baulin2007} were the first to report on a two
dimensional dynamical system of treadmilling and colliding
microtubules. Their focus was on establishing the minimal
interactions needed to generate dynamical alignment. Using
stochastic simulations they showed that a pausing mechanism,
whereby a growing microtubule stalls against another microtubule
until the latter moves away, can indeed lead to ordering.
Stalling, however, is not often observed in the cortical array.
Moreover their model lacks dynamic instabilities, i.e.
catastrophes, both spontaneous and induced, and rescues, and
employs a form of deterministic microtubule motion, which is
arguably unrealistic in view of the observed dynamics.

The outline of the paper is as follows. In section II we formulate our course-grained model starting from a
description of the dynamics of individual microtubules. We then construct the continuity equations that couple the
densities of growing, shrinking and inactive microtubule segments due to the intrinsic and collisional dynamics. In
the steady state we can reduce the initial set of equations to four coupled non-linear integral equations. We then
perform a dimensional analysis to identify the relevant control parameter of the system. In section III we
present the results of our model. We first solve the model analytically in the isotropic stationary state. Using a
bifurcation analysis we then determine the critical values of the control parameter at which the system develops
ordered solutions. We interpret these results in terms of the physical parameters of microtubule segment length and
mesh size. Finally, we formulate a minimal model with realistic interaction parameters that we can solve numerically
to obtain all stationary ordered solutions. We close by giving arguments for the stability of these solutions. The
paper concludes with a discussion section. An appendix outlines further details of the numerical solution technique
employed.

\section{Model}\label{sec:model}

\subsection{Description of the microtubules and their dynamics}
\begin{figure}[hbt]
  \begin{centering}
   \includegraphics{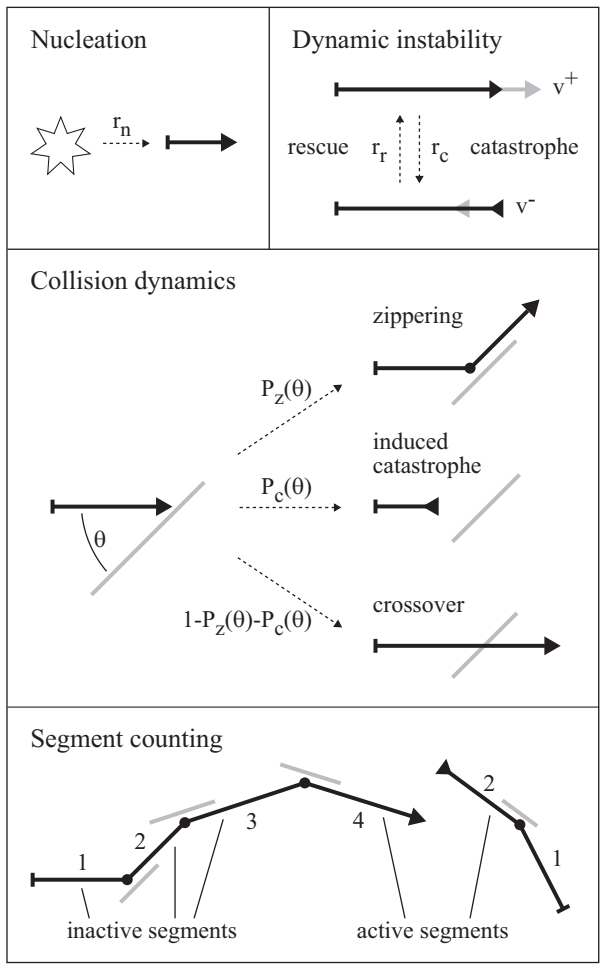}
   \caption{\label{fig:mTsegs} Schematic representation of the model interaction. The microtubule of interest is drawn in black and other microtubules that it encounters are in grey. The active segments of the black microtubule have an arrow head indicating growth or shrinkage whereas inactive segments end in the junction with the following segment depicted by a dot. See also the description of the parameters in table \ref{tab:notationDim}.}
 \end{centering}
\end{figure}

As described in the introduction we confine the configuration of the microtubules to a 2D plane. Since
collision-induced zippering events can cause microtubules to bend along the direction of preexisting ones, we divide
each microtubule into distinct segments with a fixed orientation. We treat these segments as straight rigid rods.
This is justifiable since the persistence length $l_p$ of microtubules is
long ($\sim \mathrm{mm}$) compared to the average length of a microtubule ($%
\sim 10\mu\mathrm{m}$) and, as mentioned above, adhesion to the
plasma membrane further inhibits thermal motion.

Microtubules are known to be dynamic in that they are continually
growing or shrinking by (de)polymerization. We use the standard
two-state dynamic instability model of Dogterom and Leibler
\cite{Marileen93} which assumes that each microtubule has a `plus'
end, located on the final segment of each microtubule, that is
either growing (labelled by $+$) with speed $v^{+}$ or shrinking
(labelled by $-$) with speed $v^{-}$. This plus end can switch
stochastically from growing to shrinking (a so-called
`catastrophe') with
rate $r_{\textrm{c}}$, or from shrinking to growing (a so-called `rescue') with rate $%
r_{\textrm{r}}$ in a process known as dynamic instability.

We model the creation of new microtubules with a constant,
homogeneous, isotropic nucleation rate $r_{\textrm{n}}$ in the
plane of the 2D model. \textit{In vivo} nucleation appears to
occur at the cortex and has been observed to occur in random
orientations unattached to pre-existing microtubules
\cite{Shaw+03}. Although microtubules have also been observed to
nucleate from by $\gamma $-tubulin complexes binding to
pre-existing microtubules \cite{Mur+05,Murata2007,EhrhardtShaw06}
we ignore this possibility for simplicity's sake. By the same
token we disregard the possibility of the shrinking of
microtubules at their less active \textquoteleft
minus\textquoteright\ end, leading to motion through the
`treadmilling' mechanism \cite{margolis1998}. The initial segment
of each microtubule therefore remains attached to the nucleation
point in our model.

We call the final segment of a microtubule, which contains the
growing or shrinking tip, \emph{active} and all the remaining
ones, which do not change their length, \emph{inactive} (labelled
by $0$). A cartoon of an an individual microtubule according to
these definitions is depicted in figure~\ref{fig:mTsegs}. When a
microtubule collides with another microtubule and experiences a
zippering event, its active segment is converted into an inactive
segment, and a new active segment is created alongside the
encountered microtubule. The inverse can also occur: if the active
segment shrinks to zero length, a previously inactive segment in
another direction can be reactivated. An induced catastrophe event
simply causes the growing active segment to become a shrinking
one, as is the case for spontaneous catastrophes. Finally, a
crossover results in the growing active segment continuing to grow
unperturbed.

\begin{figure}[hbt!]
  \begin{centering}
   \includegraphics{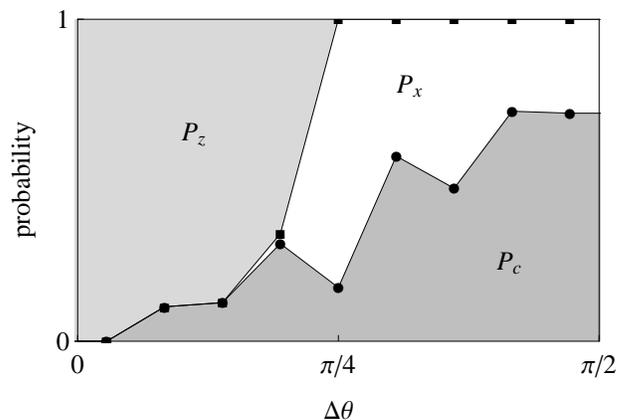}
\caption{\label{fig:int} Probabilities for zippering, cross-overs
and catastrophes as deduced from the observations of
\cite{DixCyr04} (combined data from MBD-DsRed and YFP-TUA6
labelling). Light grey shaded region: fraction of zippering
events. Dark grey shaded region: fraction of induced catastrophes.
White region: fraction of crossovers. Every data point is located
at the centre of the corresponding bin, and the shaded regions
have been extended to the boundaries using horizontal lines. The
corresponding lowest order Fourier coefficients of the interaction
functions are: $\hat{c}_0=0.59$, $\hat{c}_2=-0.36$,
$\hat{z}_0=0.24$ (computed using numerical integration of the
product of $\left| \sin{\theta} \right|$ and a piecewise linear
interpolation of the data).}
 \end{centering}
\end{figure}

In figure \ref{fig:int} we present the relative probabilities for
zippering, induced catastrophes and crossovers as a result of
collisions between microtubules, based on the data provided by
Dixit and Cyr \cite{DixCyr04}. We assume that there are no
microtubule polarity effects, as they were not reported. The
probabilities $P_{\textrm{z}}(\theta -\theta ^{\prime })$,
$P_{\textrm{x}}(\theta -\theta ^{\prime })$ and
$P_{\textrm{c}}(\theta -\theta ^{\prime })$ for zippering,
cross-overs and induced catastrophes respectively are therefore
even functions of the angle difference $\theta -\theta ^{\prime }$
defined by their values on the interval $[0,\frac{\pi}{2}]$. In
this article we will use only the following minimal set of
properties, which are qualitatively supported by the data.
Firstly, zippering becomes less likely for increasing angle of
incidence, and is effectively zero at $\theta -\theta ^{\prime
}=\frac{\pi}{2}$, which is reasonable as the energy associated
with bending the microtubule increases with angle. Secondly, the
probability for induced catastrophes monotonically increases with
increasing angle of incidence, reaching a maximum at $\theta
-\theta ^{\prime }=\frac{\pi}{2}$, consistent with observations
that indicate that a microtubule which is hindered in its growth
will undergo a catastrophe at a rate that depends inversely on its
growth speed \cite{Marileen03}.

\subsection{Continuum model}

\begin{table*}[ht]
\caption{Overview of all parameters and variables in natural dimensions}\label{tab:notationDim}
\begin{tabular}[c]{lll}\hline Parameters& Description &Dimensions\\\hline
$v^{+}$ & growth speed & $\left[ \text{length}\right]/\left[
\text{time}\right]$\\
$v^{-}$ & shrinkage speed & $\left[ \text{length}\right]/\left[
\text{time}\right]$\\
$r_{\textrm{c}}$ & catastrophe rate & $1/\left[
\text{time}\right]$\\
$r_{\textrm{r}}$ & rescue rate & $1/\left[
\text{time}\right]$\\
$r_{\textrm{n}}$ & nucleation rate & $\left[ \text{length}\right]^{-2} \left[
\text{time}\right]^{-1}$\\
$P_{\textrm{c}}(\theta)$ & probability of induced catastrophe upon
collision & 1\\
$P_{\textrm{z}}(\theta)$ & probability of zippering upon collision & 1\\
\hline Synthetic parameters & & \\\hline $g=\frac{r_{\textrm{r}}}{v^{-}}-\frac{r_{\textrm{c}}}{v^{+}}$ & growth
parameter &
$1/\left[ \text{time}\right]$\\
$u=1+\frac{v^{+}}{v^{-}}$ & speed ratio & 1\\
$c(\theta)=\sinb{|\theta|}P_{\textrm{c}}(\theta)\longleftrightarrow\left\{
\hat{c}_{n}\right\} $ & effective catastrophic collision probability & 1\\
$z(\theta)=\sinb{|\theta|}P_{\textrm{z}}(\theta)\longleftrightarrow\left\{
\hat{z}_{n}\right\} $ & effective zippering probability &
1\\\hline Dependent variables & & \\\hline $k\left( \theta\right)$
& microtubule length density & $\left[ \text{length}\right]^{-1}
\left[
\text{radian}\right]^{-1}$\\
$l\left(  \theta\right)  $ &
average microtubule segment length & $\left[ \text{length}\right]$\\
$\left\{m_{i}^{+}\left(l, \theta\right) ,m_{i}^{-}\left(
l,\theta\right),m_{i}^{0}\left(l, \theta\right) \right\}  $ &
density of growing/shrinking/inactive segments & \\
& with length $l$ and direction $\theta$ & $\left[ \text{length}\right] ^{-3}\left[ \text{radian}\right] ^{-1}$\\
\hline
\end{tabular}
\end{table*}

Since there are many ($\approx 10^{2}$ - $10^{3}$) microtubules,
each of which can have multiple segments, in the cortical array of
a typical interphase plant cell we treat the system using a
coarse-grained description. In this approach, instead of
individual microtubules, we consider local densities of
microtubule segments. This approximation is reasonable as long as
the length scale of an individual microtubule segment is small
compared to the linear dimensions of the cell. From the outset we
assume that the system is (and remains) spatially homogeneous, and
we will eventually restrict ourselves to the steady-state
solutions. In order to deal with the memory effect caused by the
isotropic nucleation, followed by subsequent reorienting zippering
events, we need to keep track of the segment number $i$, which
starts at $1$ for the segments connected to their nucleation site
and increases by unity at each zippering event. Our fundamental
variables are therefore the areal number densities $m_{i}^{\sigma
}(l,\theta ,t)$ of segments in state $\sigma \in \left\{
0,-,+\right\} $ with segment number $i$ having length $l$ and
orientation $\theta $ (measured from an arbitrary axis) at time
$t.$ These densities obey a set of master equations that can
symbolically be written as
\begin{subequations}\label{eq:MEall}
\begin{align}
\partial _{t}m_{i}^{+}(l_{i},\theta _{i},t) &=\Phi _{\text{\textit{growth}}%
}+\Phi _{\text{\textit{rescue}}}-\Phi _{\text{\textit{spont. cat.}}}\nonumber\\&-\Phi _{%
\text{\textit{induced cat.}}}-\Phi _{\text{\textit{zipper}}}
\label{eq:MEgrow} \\
\partial _{t}m_{i}^{-}(l_{i},\theta _{i},t)& =\Phi _{\text{\textit{shrinkage%
}}}-\Phi _{\text{\textit{rescue}}}+\Phi _{\text{\textit{spont.
cat.}}}\nonumber\\&+\Phi _{\text{\textit{induced cat.}}}+\Phi
_{\text{\textit{reactivation}}}
\label{eq:MEshrink} \\
\partial _{t}m_{i}^{0}(l_{i},\theta _{i},t)& =+\Phi _{\text{\textit{zipper}}%
}-\Phi _{\text{\textit{reactivation}}}  \label{eq:MEinactive}
\end{align}%
\end{subequations}
The flux terms $\Phi _{\text{\textit{event}}}$ couple the
equations for the growing, shrinking and inactive segments and
between different values of $i$. Equations \eqref{eq:MEall} must
be supplemented by a set of boundary conditions for the growing
segments at $l=0$. For the initial segment ($i=1$) this reflects
the isotropic nucleation of new microtubules, given by
\begin{equation}
v^{+}m_{1}^{+}(l_{1}=0,\theta ,t) =\frac{r_{\textrm{n}}}{2\pi},
\label{eq:nuc1}
\end{equation}
where $r_{\textrm{n}}$ is nucleation rate. For the subsequent
segments $i>1$, this `nucleation' of growing segments is the
result of the zippering of segments with index $i-1$. Defining
$\varphi _{\text{zipper}}\left( \theta _{i-1}\rightarrow \theta
_{i},l_{i-1},t\right)$ as the flux of $i$-segments with angle
$\theta_i$ and length $l_i$ zippering into angle $\theta_{i+1}$ at
time $t$ (this will be made explicit in equation
\eqref{eq:defPhiZipper}), we obtain the boundary condition
\begin{align}
v^{+}m_{i\geq 2}^{+}(l_{i} =0,\theta _{i},t)&=\nonumber\\\int
\mathrm{d}l_{i-1}\int \mathrm{d}\theta _{i-1}&\;\varphi
_{\text{zipper}}\left( \theta _{i-1}\rightarrow \theta
_{i},l_{i-1},t\right) \;  \label{eq:nuci}.
\end{align}
Generally, this leads to a qualitatively different boundary
condition for every value of $i$. The model therefore consists of
an infinite set of coupled equations, three for every value of
$i$. However, in section \ref{sec:steady} we will show that in the
steady state, this can be reduced to a finite set by summing over
all segment indices $i$. In the following, we derive explicit
expressions for each of the flux terms $\Phi
_{\text{\textit{event}}}$.

\subsubsection{Growth and shrinkage terms: $\Phi _{\text{growth}}$, $\Phi _{\text{shrinkage}}$}

$\Phi _{\text{\textit{growth}}}$ in Equation \eqref{eq:MEgrow}
corresponds to the length increase of the growing segments. For
segment growth in isolation, the length increase in a small time
interval $\delta t$ is given by $v^{+}\delta t,$ where $v^{+}$ is
the growth velocity, and we have $m^{+}(l+v^{+}\delta t,\theta
,t+\delta t)=m^{+}(l,\theta ,t)$. By expanding the left hand term to first order in $\delta t,$ we find%
\begin{equation}
\partial _{t}m _{i}^{+}(l,\theta ,t)=-v^{+}\partial _{l}m_{i}^{+}(l,\theta
,t)\equiv \Phi _{\text{\textit{growth}}}  \label{eq:PHIgrow}
\end{equation}%
A similar derivation yields that
\begin{equation}
\partial _{t}m _{i}^{-}(l,\theta ,t)=v^{-}\partial _{l}m_{i}^{-}(l,\theta
,t)\equiv \Phi _{\text{\textit{shrink}}}  \label{eq:PHIshrink}
\end{equation}
where $v^{-}$ is the shrinking velocity.

\subsubsection{Dynamic instability terms: $\Phi _{\text{rescue}}$ ,$\Phi _{%
\text{spont. cat.}}$}

$\Phi_{\text{\textit{rescue}}}$ and $\Phi_{\text{\textit{spont. cat.}}}$ in equations~\eqref{eq:MEgrow} and
\eqref{eq:MEshrink} correspond to the fluxes due to the spontaneous rescues and spontaneous catastrophe respectively
and are simply given by
\begin{align}
\Phi _{\text{\textit{rescue}}}=& r_{\textrm{r}} m_{i}^{-}(l_{i},\theta _{i},t)
\label{eq:PHIrescue} \\
\Phi _{\text{\textit{spont. cat.}}}=& r_{\textrm{c}} m_{i}^{+}(l_{i},\theta _{i},t) \label{eq:PHIspont}
\end{align}%
where $r_{\textrm{r}}$ is the spontaneous rescue rate and $r_{\textrm{c}}$ is the spontaneous catastrophe rate.

So far, we have described the first three terms of equations
\eqref{eq:MEgrow} and \eqref{eq:MEshrink} (growth, shrinkage and
dynamic instability terms). Together, these fully describe a
system of non-interacting microtubules, in which also the boundary
condition \eqref{eq:nuci} vanishes due to the absence of
zippering. In this special case we recover the well-known
equations introduced by Dogterom and Leibler \cite{Marileen93}
(for $i=1$).

\subsubsection{Interaction terms: $\Phi _{\text{induced cat.}}$, $\Phi _{\text{%
zipper}}$}

An interaction can occur when a growing active microtubule segment collides with another segment, irrespective of
the latter's state and length. This prompts the definition of the total length density $k(\theta ,t)$ of all
microtubule segments in direction $\theta $ at time $t$, given by:
\begin{equation}
k(\theta ,t)=\sum_{i}\int \mathrm{d}l_{i}\,l_{i}(m_{i}^{+}(l_{i},\theta ,t)+m_{i}^{-}(l_{i},\theta
,t)+m_{i}^{0}(l_{i},\theta ,t))\; \label{eq:ldensdef}
\end{equation}
The density of collisions of a microtubule segment growing in
direction $\theta $ with other segments in direction
$\theta^{\prime}$ is determined by the geometrical projection
\begin{equation}
\left\vert \sinb{ \theta -\theta ^{\prime }} \right\vert k(\theta
^{\prime },t),
\end{equation}
where the presence of the $\sinb{ \theta -\theta ^{\prime }}$ factor ensures the correct geometrical weighting reflecting
the fact that parallel segments do not collide.
When a collision occurs, one of the three possible events, induced
catastrophe $\left( c\right) ,$ zippering $\left( z\right) $ or cross-over $%
\left( x\right) $ occurs, with probabilities $P_{\textrm{c}}\left( \theta -\theta
^{\prime }\right) ,$ $P_{\textrm{z}}\left( \theta -\theta ^{\prime }\right) $ and $%
P_{\textrm{x}}\left( \theta -\theta ^{\prime }\right) $
respectively. These probabilities can (and \emph{in-vivo} do, see
Figure \ref{fig:int}) depend on the relative angle $\theta -\theta
^{\prime }$ between the incoming segment and the \ \textquoteleft
scatterer\textquoteright. For convenience sake we absorb the
geometrical factor $\left\vert \sinb{ \theta -\theta
^{\prime }} \right\vert $ into the probabilities, by defining $%
f\left( \theta -\theta ^{\prime }\right) =\left\vert \sinb{
 \theta -\theta ^{\prime }} \right\vert P_{f}\left(
\theta -\theta ^{\prime }\right) $ for all events $f\in \left\{ c,z,x\right\} .$ The incoming flux of growing
microtubule segments with given segment number, length and orientation is given by $v^{+}m_{i}^{+}(l,\theta ,t).$
With these definitions we can write
the interaction terms as%
\begin{align}
\Phi _{\text{induced cat.}} &=v^{+}m_{i}^{+}(l_{i},\theta _{i},t)\int \mathrm{d}\theta ^{\prime }\,c(\theta
_{i}-\theta ^{\prime })k(\theta ^{\prime
},t) \;  \label{eq:PHIind} \\
\Phi _{\text{zipper}} &=v^{+}m_{i}^{+}(l_{i},\theta _{i},t)\int \mathrm{d}%
\theta ^{\prime }\,z(\theta _{i}-\theta ^{\prime })k(\theta ^{\prime },t) \;  \label{eq:PHIzip}
\end{align}
The analogous term for crossovers is not used, because the occurrence of a crossover event has no effect on the
growth of a microtubule.

\subsubsection{Reactivation term: $\Phi _{\text{reactivation}}$}

The reactivation term $\Phi _{\text{reactivation}}$ corresponds to
the flux of active microtubule segments, with segment number
$i+1,$ that, by shrinking to length zero, reactivate a previously
inactive segment, effectively undoing a past zippering event and
so create a new, shrinking, active segment with segment number
$i.$ The incoming flux of of such segments coming from a
given direction $\theta _{i+1}$ is given by $v^{-}m_{i+1}^{-}(l_{i+1}=0,%
\theta _{i+1},t)$. The reactivation flux is given by
\begin{multline}
\Phi _{\text{reactivation}}=\\\int d\theta
_{i+1}\,v^{-}m_{i+1}^{-}(l_{i+1}=0,\theta
_{i+1},t)p_{\text{unzip}}(\theta _{i},l_{i}|\theta_{i+1}, t),
\label{eq:PHIreact}
\end{multline}
where the \textquoteleft unzippering\textquoteright\ distribution\
$p _{\text{unzip}}(\theta _{i},l_{i}|\theta_{i+1}, t)$ gives the
probability that the shrinking microtubule reactivates an inactive
segment with orientation $\theta _{i}$ and length $l_{i}$. This
distribution will be determined below.

A microtubule that has zippered will take a certain amount of time
$\tau$ to undergo a catastrophe and return to the zippering
location, where $\tau$ is a stochastic variable. The unzipppering
flux from direction $\theta_{i+1}$ at time $t$ consists of
microtubules that had zippered at a range of times $t - \tau$ and
have now returned to the zippering location. This implicitly
defines an \emph{originating time} distribution
$p_{\text{origin}}\left( t-\tau |\theta _{i+1},t \right)$ for the
returning microtubules. Furthermore, because the evolution of a
microtubule between the zippering event and its return to the same
location does not depend on the previous segments, the segment
that is re-activated by a microtubule returning to the zippering
position after a time $\tau$ should be selected proportional to
the `forward' zippering flux at time $t-\tau$. The forward flux
$\varphi _{\text{zipper}}\left( \theta _{i}\rightarrow \theta
_{i+1},l_{i},t\right)$ of microtubules with length $l_i$ and angle
$\theta_i$ zippering into angle $\theta_{i+1}$ is defined in
accordance with equation \eqref{eq:PHIzip} as
\begin{multline}
\varphi _{\text{zipper}}\left( \theta _{i}\rightarrow \theta
_{i+1},l_{i},t\right) =\\v^{+}m_{i}^{+}(l_{i},\theta
_{i},t)z(\theta _{i}-\theta _{i+1})k(\theta _{i+1},t).
\label{eq:defPhiZipper}
\end{multline}
At each of the originating times $t - \tau$, the distribution of
microtubules that zipper into the direction $\theta_{i+1}$ with
length $l_{i}$ and orientation $\theta _{i}$ is given by
\begin{equation}
p_{\text{zip}}(\theta _{i},l_{i}|\theta_{i+1},
t-\tau)=\frac{\varphi _{\text{zipper}}\left( \theta
_{i}\rightarrow \theta _{i+1},l_{i},t-\tau \right) }{\int
\mathrm{d}l^{\prime }\mathrm{d}\theta ^{\prime }\;\varphi
_{\text{zipper}}(\theta ^{\prime }\rightarrow \theta
_{i+1},l^{\prime },t-\tau )\;}.
\end{equation}
The probability distributions $p_{\text{origin}}\left( t-\tau
|\theta _{i+1},t \right)$ and $p_{\text{zip}}(\theta
_{i},l_{i}|\theta_{i+1}, t-\tau)$ can be combined to determine the
unzippering distribution
\begin{multline}
p_{\text{unzip}}(\theta _{i},l_{i}|\theta_{i+1},
t)=\int_{0}^{t}\!\mathrm{d}\tau \,p_{\text{origin}}\left( t - \tau
|\theta _{i+1},t\right) \times
\\ \frac{\varphi _{\text{zipper}}\left( \theta
_{i}\rightarrow \theta _{i+1},l_{i},t-\tau \right) }{\int \mathrm{d}%
l^{\prime }\mathrm{d}\theta ^{\prime }\;\varphi
_{\text{zipper}}(\theta ^{\prime }\rightarrow \theta
_{i+1},l^{\prime },t-\tau )\;},
\end{multline}%
where we assume the system evolved from an initial condition at
$t_{0}=0$ in which no microtubules were present. Clearly all the
complicated history dependence of the system is hidden in the
originating time distribution. However, in the steady state situation
we consider below, the time-dependence drops out and the details
of this distribution become irrelevant.

\subsection{The steady state}

\label{sec:steady} We now consider the steady state of the system
of equations we have formulated. Setting the time derivatives to
zero, the sum of equations \eqref{eq:MEgrow} to
\eqref{eq:MEinactive} yields $\Phi _{\text{\textit{growth}}}+\Phi
_{\text{\textit{shrinkage}}}=0$, which together with equation
\eqref{eq:PHIgrow} and \eqref{eq:PHIshrink} implies $\partial
_{l_{i}}\left[ v^{+}m_{i}^{+}(l_{i},\theta
_{i})-v^{-}m_{i}^{-}(l_{i},\theta _{i})\right] =0$. Because
physically acceptable solutions should be bounded as
$l_{i}\rightarrow \infty ,$ we obtain the length flux balance
equation
\begin{equation}
v^{+}m_{i}^{+}(l_{i},\theta _{i})=v^{-}m_{i}^{-}(l_{i},\theta _{i}), \label{eq:balance}
\end{equation}%
showing that the growing and shrinking segments have, up to a constant amplitude, the same orientational and length
distribution. This allows us to eliminate $m_{i}^{-}(l_{i},\theta _{i})$ from \eqref{eq:MEgrow} to obtain
\begin{multline}
\partial _{l_{i}}m_{i}^{+}(l_{i},\theta _{i})= m_{i}^{+}(l_{i},\theta _{i})  \times\\
\left\{ g-\int \mathrm{d}\theta ^{\prime }\;\left[c(|\theta _{i}-\theta ^{\prime }|)+z(|\theta _{i}-\theta ^{\prime
}|)\right]k(\theta ^{\prime }) \right\}, \label{eq:SSgrow}
\end{multline}
where the \emph{growth parameter }
\begin{equation}
g=\frac{r_{\textrm{r}}}{v^{-}}-\frac{r_{\textrm{c}}}{v^{+}}.
\end{equation}%
characterizes the behaviour of the bare, non-interacting, system
in which microtubules remain bounded in length for $g<0$ and
become unbounded for $g\geq 0.$ As the bracketed factor on the
right hand side of \eqref{eq:SSgrow} does not depend on the
segment length nor on the segment number, we immediately obtain
that $m_{i}^{+}(l_{i},\theta _{i})$ has an exponential length
distribution
\begin{equation}
m_{i}^{+}(l_{i},\theta _{i})=m_{i}^{+}(\theta
_{i})e^{-l_{i}/l(\theta _{i})} \label{eq:expl}
\end{equation}%
where the average segment length $l\left( \theta_i \right) $ in the direction $%
\theta_i $ is given by
\begin{equation}
\frac{1}{l(\theta )}=-g+\int \mathrm{d}\theta ^{\prime }\,(c(\theta -\theta ^{\prime })+z(\theta -\theta ^{\prime
}))k(\theta ^{\prime }). \label{eq:Dcoll}
\end{equation}
The nucleation boundary conditions \eqref{eq:nuc1} and
\eqref{eq:nuci} are now transformed into independent nucleation
equations that are expressed in terms of the amplitudes
$m_{i}^{+}(\theta _{i})$
\begin{align}
v^{+}m_{1}^{+}(\theta )& =\frac{r_{\textrm{n}}}{2\pi},  \label{eq:SSnuc1} \\
m_{i\geq 2}^{+}(\theta )& =k(\theta )\int \mathrm{d}\theta
^{\prime }\;z(\theta ^{\prime }-\theta )l(\theta ^{\prime
})m_{i-1}^{+}(\theta ^{\prime }).  \label{eq:SSnuci}
\end{align}

We now note that under these conditions equation
\eqref{eq:MEinactive} is already satisfied, as can be explicitly
checked by considering $\Phi _{\text{reactivation}}$
\eqref{eq:PHIreact} and using that in the steady state $\varphi
_{\text{zipper}}$ does not depend on time and the integral over
$p_{\text{origin}}$ is by definition equal to 1. This, in
combination with the results (\ref{eq:balance}), (\ref{eq:expl})
and (\ref{eq:SSnuci}), yields the identity with $\Phi
_{\text{zipper}}.$ We therefore need an independent argument to
fix the densities of the inactive segments. To obtain this we use
the steady state rule that \textsf{population size} $=$
\textsf{nucleation rate} $\times $ \textsf{average lifetime}.
Consider a newly \textquoteleft born\textquoteright\ growing
segment, created either by a nucleation or a zippering event. Its
average life time is by definition the average time until it
shrinks back to zero length, i.e. the average return time. Clearly
this time only depends on its orientation $\theta $, the steady
state microtubule length density $k\left( \theta ^{\prime }\right)
$ and the dynamical instability parameters, but not on the segment
number. We therefore denote it by $\tau \left( \theta \right)$.
The steady state density of inactive segments with length $l_{i}$,
orientation $\theta $ and segment number $i$ is then given by
\begin{equation}
m_{i}^{0}\left( l_{i},\theta \right) =\int \mathrm{d}\theta
^{\prime }\,\!\,\varphi _{\text{zipper}}\left( \theta \rightarrow
\theta ^{\prime },l_{i}\right) \tau \left( \theta ^{\prime
}\right), \label{eq:SSinact}
\end{equation}
where $\varphi _{\text{zipper}}$ is defined by equation
\eqref{eq:defPhiZipper}, as inactive segments are created by a
zippering event. Because the only length-dependent term on the
right-hand side is $m^+(l_i,\theta)$, it follows that the
length-dependence of the inactive segment distributions is
proportional to those of the active segments, i.e.
\begin{equation}
m_{i}^{0}\left( l_{i},\theta \right) =m_{i}^{0}\left( \theta \right) e^{-%
\frac{l_{i}}{l\left( \theta \right) }}.
\end{equation}

At the same time, the total integrated length density of segments,
both active and inactive, with segment number $i+1$ in the
direction $\theta ^{\prime }$ is given by
\begin{multline}
N_{i+1}^{\text{total}}\left( \theta ^{\prime }\right) =\int
\mathrm{d} l_{i+1}\, \times  \\\left[ m_{i+1}^{0}\left(
l_{i+1},\theta ^{\prime }\right) +m_{i+1}^{+}\left( l_{i+1},\theta
^{\prime }\right)
+m_{i+1}^{-}\left( l_{i+1},\theta ^{\prime }\right) \right]\\
 =\int \mathrm{d}\theta ^{\prime \prime }\,\!\int
\mathrm{d}l_{i}\,\varphi _{\text{zipper}}\left( \theta ^{\prime
\prime }\rightarrow \theta ^{\prime },l_{i}\right) \tau \left(
\theta ^{\prime }\right),
\end{multline}
where the last equality follows from the fact that every segment
with index $i+1$ has been created by a zippering event of a
segment with index $i$. We solve this for $\tau \left( \theta
^{\prime }\right)$ and insert the result in (\ref{eq:SSinact}),
which, after expanding $\varphi_{\text{zipper}}$
\eqref{eq:defPhiZipper}, produces the following expression for
$m_{i}^{0}\left( \theta \right)$:
\begin{multline}
m_{i}^{0}\left( \theta \right) =m_{i}^{+}(\theta )\int
\mathrm{d}\theta
^{\prime }\,\!\,z(\theta -\theta ^{\prime })l(\theta ^{\prime })\times \\
\frac{ \left[ m_{i+1}^{0}\left( \theta ^{\prime }\right)
+m_{i+1}^{+}\left( \theta ^{\prime }\right) +m_{i+1}^{-}\left(
\theta ^{\prime }\right) \right] }{ \int \mathrm{d}\theta ^{\prime
\prime }\,\!z(\theta ^{\prime \prime }-\theta ^{\prime })l\left(
\theta ^{\prime \prime }\right) m_{i}^{+}(\theta ^{\prime \prime
})}. \label{eq:m0definition}
\end{multline}
We use the nucleation equation (\ref{eq:SSnuci}) to replace the
integral in the denominator of the integrand on the right hand
side of this expression. In addition, we define the quantity
$Q_i(\theta)$ through
\begin{align}
m_{i}^{0}\left( \theta \right) &=Q_{i}\left( \theta \right) \left[ m_{i}^{+}\left( \theta \right) +m_{i}^{-}\left(
\theta \right)
\right] \nonumber \\
&=\left( 1+\frac{v^{+}}{v^{-}}\right) Q_{i}\left( \theta \right)
m_{i}^{+}\left( \theta \right) \nonumber\\
&\equiv u Q_{i}\left( \theta \right) m_{i}^{+}\left( \theta
\right)\label{eq:defQ}
\end{align}
and equation \eqref{eq:m0definition} leads to the following
recursion relation for $Q_{i}\left( \theta \right)$
\begin{equation}
Q_{i}\left( \theta \right) =\int \mathrm{d}\theta ^{\prime
}\;z\left( \theta -\theta ^{\prime }\right) k\left( \theta
^{\prime }\right) l\left( \theta ^{\prime }\right) \left(
1+Q_{i+1}\left( \theta ^{\prime }\right) \right) .
\label{eq:SSconnecti}
\end{equation}
We now argue that the ratio $Q_{i}(\theta )$ is in fact
\emph{in}dependent of the segment number. Using the fact that the
growing, shrinking and inactive segments have an identical
exponential profile, it follows from \eqref{eq:defQ} that
$Q_{i}(\theta )$ is equal to the ratio between inactive and active
segments
\begin{equation}
Q_{i}(\theta )=\frac{m_{i}^{0}\left( \theta \right)
}{um_{i}^{+}\left( \theta \right) }=\frac{N_{i}^{0}\left( \theta
\right) }{N_{i}^{+}\left( \theta \right) +N_{i}^{-}\left( \theta
\right) }.
\end{equation}
After a new microtubule segment has been created it will generally
spend some time in an active state and some time time in an
inactive state. The expected lifetime $\tau(\theta)$ can also be
separated into the expected active and inactive lifetimes for any
newly created segment: $\tau \left( \theta \right) =\tau
_{\text{active}}\left( \theta \right) +\tau
_{\text{inactive}}\left( \theta \right) $. These lifetimes are
necessarily proportional to the total number of active and
inactive segments, so that $Q_i(\theta) = \tau
_{\text{inactive}}\left( \theta \right) / \tau
_{\text{active}}\left( \theta \right)$. As we have argued before,
these lifetimes do not depend on the segment number, and, hence,
neither does $Q_{i}(\theta )$. An alternative route to the same
conclusion follows from expanding out the forward recursion in
(\ref{eq:SSconnecti}) to show that $Q_{i}\left( \theta \right) $
can for every $i$ formally be written as the same infinite series
of multiple integrals involving $z\left( \theta -\theta ^{\prime
}\right)$, $k\left( \theta \right)$ and $l\left( \theta \right)$.
We therefore write the self-consistency relationship
\begin{equation}
Q\left( \theta \right) =\int \mathrm{d}\theta ^{\prime }\;z\left(
\theta -\theta ^{\prime }\right) k\left( \theta ^{\prime }\right)
l\left( \theta ^{\prime }\right) \left( 1+Q\left( \theta ^{\prime
}\right) \right). \label{eq:SSconnect}
\end{equation}

The final closure of this set of equations is provided by the definition of the length density (\ref{eq:ldensdef})
applied to the steady state
\begin{align}
k(\theta ) &=\sum_{i}\int \mathrm{d}l_{i}\,l_{i}\left[ m_{i}^{+}(l_{i},\theta )+m_{i}^{-}(l_{i},\theta
)+m_{i}^{0}(l_{i},\theta
)\right] \; \nonumber \\
&=ul(\theta )^{2}(1+Q(\theta ))\sum_{i}m_{i}^{+}(\theta ).
\label{eq:SSdens}
\end{align}

\subsection{Dimensional analysis}

\label{sec:dimension} In order to simplify our equations for further analysis and to identify the relevant control
parameter we perform a dimensional analysis. We therefore introduce a common length scale and rescale all lengths
with respect to this length scale. For example, our primary variables $m_{i}^{+}(\theta )$ have dimension $\left[
\text{length}\right] ^{-3}\left[ \text{radian}\right] ^{-1}$. Taking our cue from (\ref{eq:SSnuc1}) and
\eqref{eq:SSdens} we adopt the length scale
\begin{equation}
l_{0}=\left( \frac{1}{\pi} \frac{v^{+}}{u \frac{r_{\textrm{n}}}{2 \pi}}\right) ^{\frac{1}{3}}, \label{eq:l0def}
\end{equation}
where the additional factor of $\pi^{-1}$ within the parentheses is added to suppress explicit factors involving
$\pi$ in the final equations. This definition allows us to define the dimensionless variables
\begin{subequations}\label{eq:dimensionlessTransform}
\begin{align}
L\left( \theta \right) & =l\left( \theta \right) /l_{0} \\
K\left( \theta \right) & =\pi k\left( \theta \right)l_0  \\
M_{i}^{+}\left( \theta \right) & =\pi m_{i}^{+}\left( \theta \right)l_{0}^{3} \label{eq:mPlusDimensionless} \\
G& =g l_{0}.
\end{align}
In the absence of interactions, \eqref{eq:Dcoll} shows that the
average length $l$ of the microtubule is given by $l=-1/g$. This
implies $G=-l_0/l$, meaning that, for $G<0$, $G$ can be
interpreted as a measure for the non-interacting microtubule
length.

In addition, we adopt the dimensionless operator notation%
\begin{equation}
\mathbf{F}\left[ h\right] \left( \theta \right) =\frac{1}{\pi}\int_{0}^{2\pi }\mathrm{d}%
\theta ^{\prime }\;f\left( \theta -\theta ^{\prime }\right) h\left( \theta ^{\prime }\right), \label{eq:operatorDef}
\end{equation}
\end{subequations}
where $F\in \left\{ C,Z\right\}$. We are now in a position to
express the equations in terms of the dimensionless quantities.
Applying \eqref{eq:mPlusDimensionless} to the nucleation equations
\eqref{eq:SSnuc1} and \eqref{eq:SSnuci} yields expressions for the
growing segment densities $M_i^+(\theta)$. Furthermore, the
$M_i^+(\theta)$ for the different segment labels can be absorbed
into a single microtubule plus end density (density of active
segments), given by
\begin{equation}
T\left( \theta \right) = u L\left( \theta \right)
\sum_{i=1}^{\infty }M_{i}^{+}\left( \theta \right).
\end{equation}
Performing all substitutions, the final set of dimensionless
equations reads

\bigskip
\newlength{\mylength}
\setlength{\fboxsep}{10pt} \setlength{\mylength}{\linewidth} \addtolength{\mylength}{-2\fboxsep}
\addtolength{\mylength}{-2\fboxrule}
\noindent
\fbox{%
\parbox{\mylength}{ \setlength{\abovedisplayskip}{0pt} \setlength{\belowdisplayskip}{0pt}
\begin{subequations}\label{eq:setDimensionless}
\begin{align}
&\quad\textrm{Segment length} \nonumber\\
\frac{1}{L\left( \theta \right) }&= -G+\mathbf{C}\left[ K\right] \left( \theta \right) +\mathbf{Z}\left[ K\right]
\left( \theta
\right) \label{eq:coll} \\
& \quad\textrm{Density}\nonumber\\
K(\theta) &= L(\theta)(1+Q(\theta))T(\theta)\\
& \quad\textrm{Inactive-active ratio}\nonumber\\
Q\left( \theta \right) &=\mathbf{Z}\left[ LK\left( 1+Q\right) \right] \left( \theta \right) \label{eq:connectA} \\
& \quad\textrm{Plus end
density}\nonumber\\
T\left( \theta \right) &= L\left( \theta \right) +L\left( \theta \right) K\left( \theta \right) \mathbf{Z}\left[
T\right] \left( \theta \right) \label{eq:nucl}\\
\intertext{with} G &= \left[ \frac{2 v^+ v^-}{r_{\textrm{n}}\left( v^+ + v^- \right)}\right] ^{\frac{1}{3}}\left(
\frac{r_{\textrm{r}}}{v^{-}}-\frac{r_{\textrm{c}}}{v^{+}} \right) \label{eq:Gdefinition}
\end{align}
\end{subequations}
}}
\bigskip

Looking at the resulting equations, we see that the segment length
$L$ is determined by the intrinsic growth dynamics ($G$) and the
collisions leading to induced catastrophes and zippering. The
segment length density $K$ is the product of the plus end density,
the ratio of all segments to active segments ($1+Q$) and the
average segment length. The ratio $Q$ of inactive to active
segments is modulated by the zippering operator, and the plus end
density $T$ consists of contributions from direct nucleation and
zippered segments. We only consider parameter regions with
physically realizable solutions that have have real and positive
values for $L$, $K$, $Q$ and $T$.

Finally, we note that the interaction operators defined by
\eqref{eq:operatorDef} are convolutions of the operand with the
interaction functions $c(\theta)$ and $z(\theta)$. Both
interaction functions are symmetric and $\pi$-periodic, and can
therefore be written in terms of their Fourier coefficients as
\begin{align}
f(\theta) &= \frac{\hat{f}_0}{2} +
\sum_{n=1}^{\infty}\hat{f}_{2n}\cosb{2n \theta} \\
\hat{f}_{2n}&=\frac{1}{\pi}
\integral{f(\theta)\cosb{2n\theta}}{\theta}{0}{2\pi}
\end{align}
Using the identity $ \cosb{\theta-\theta'} =
\cosb{\theta}\cosb{\theta'} + \sinb{\theta}\sinb{\theta'} $ we
find that the functions $\cosb{2n\theta}$ and $\sinb{2n\theta}$
are eigenfunctions of the operators $\mathbf{C}$ and $\mathbf{Z}$,
with the Fourier coefficients $\hat{c}_{2n}$ and $\hat{z}_{2n}$,
respectively, as eigenvalues:
\begin{equation}
\mathbf{F}\left[ \cosb{2n\theta} \right] =
\hat{f}_{2n}\cosb{2n\theta} \label{eq:fouriereigenvalue}
\end{equation}
This convenient property will be exploited in later sections.

\section{Results}

\subsection{Isotropic solution}

In the isotropic phase all angular dependence drops out. Because $\textbf{C}[1] = \hat{c}_0$ and $\textbf{Z}[1] =
\hat{z}_0$ we are left with the set of equations
\begin{subequations}
\begin{align}
\frac{1}{\bar{L}}& =-G+\left( \hat{c}_{0}+\hat{z}_{0}\right) \bar{K} \label{eq:isoEqn1}\\
\bar{K}& =\bar{L}(1+\bar{Q})\bar{T} \label{eq:isoEqn2}\\
\bar{Q}& = \hat{z}_0 \bar{L}\bar{K}(1+\bar{Q})\\
\bar{T}& =\bar{L}+\hat{z}_{0}\bar{L}\bar{K}\bar{T}
\end{align}
\end{subequations}
where the overbar denotes quantities evaluated in the isotropic
phase. Solving for $\bar{Q}$ and $\bar{T}$ and inserting this into
equation \eqref{eq:isoEqn2} readily gives
\begin{equation}
\bar{K} = \frac{\bar{L}^2}{(1-\hat{z}_0 \bar{L}\bar{K})^2},
\label{eq:isoIntermediate1}
\end{equation}
which can be combined with equation \eqref{eq:isoEqn1} to yield the following relationship between $G$ and the
density
\begin{equation}
\bar{K}\left( \hat{c}_{0}\bar{K}-G\right) ^{2}=1. \label{eq:isoDensity}
\end{equation}
We see that the isotropic density is an increasing function of the
microtubule dynamics parameter $G$ and does not depend on the
amount of zippering. This can be understood by the fact that
zippering only serves to reorient the microtubules, which has no
net effect in the isotropic state. In the absence of induced
catastrophes ($\hat{c}_0=0$), the density $\bar{K}$ diverges as $G \uparrow
0$, consistent with the result by Dogterom and Leibler
\cite{Marileen93}. In the presence of induced catastrophes a
stationary isotropic solution exists for all values of $G$,
although this solution need not actually be stable.

\subsection{Bifurcation analysis}

We now search for a bifurcation point by considering the existence of steady-state solutions which are small
perturbations away from the isotropic solution. These solutions are parametrized as follows
\begin{subequations}
\begin{align}
L & =\bar{L}\left( 1+\lambda\right) \\
K & =\bar{K}\left( 1+\kappa\right) \\
Q & =\bar{Q}\left( 1+\chi\right) \\
T & =\bar{T}\left( 1+\tau\right)
\end{align}
\end{subequations}
Inserting these expressions into \eqref{eq:setDimensionless},
subtracting the isotropic solutions and expanding to first order
in the perturbations gives
\begin{subequations}
\begin{align}
\lambda &= -\bar{N}\left( \mathbf{C}[\kappa] + \mathbf{Z}[\kappa] \right) \label{eq:bifeqn1}\\
\kappa &= \lambda + \tau + \hat{z}_0 \bar{N} \chi \label{eq:bifeqn2}\\
\chi &= \frac{1}{\hat{z}_0} \mathbf{Z}\left[\lambda + \kappa + \hat{z}_0 \bar{N} \chi \right] \label{eq:bifeqn3}\\
\tau &= \lambda + \bar{N}(\hat{z}_0 \kappa + \mathbf{Z}[\tau]), \label{eq:bifeqn4}
\end{align}
\end{subequations}
where $\bar{N}=\bar{L}\bar{K}$. Note that in these equations, $\bar{N}$ has become the control parameter instead of
$G$. Using \eqref{eq:bifeqn2} and exploiting the linearity of $\mathbf{Z}$, we expand
\begin{align}
\mathbf{Z}\left[ \kappa \right] &= \mathbf{Z}\left[ \tau \right] +
\mathbf{Z}\left[ \lambda + \kappa + \hat{z}_0 \bar{N} \chi \right]
-
\mathbf{Z}\left[ \kappa \right] \\
&= \frac{1}{\bar{N}}(\tau - \lambda)-\hat{z}_0 \kappa + \hat{z}_0 \chi -\mathbf{Z}\left[ \kappa \right].
\end{align}
Solving this for $\mathbf{Z}\left[ \kappa \right]$ and inserting
the result into equation \eqref{eq:bifeqn1}, combined with
\eqref{eq:bifeqn2}, yields the relation
\begin{equation}
(1-\hat{z}_0 \bar{N})\kappa = -2 \bar{N} \mathbf{C}[\kappa ] \label{eq:bifEquation}
\end{equation}

In the \emph{absence} of induced catastrophes
($\mathbf{C}[\kappa]=0$; only zippering), a bifurcation can only
occur if $\hat{z}_0 \bar{N}=1$, which only happens for diverging density and $G=0$ (as seen from equations (\ref{eq:isoIntermediate1}-\ref{eq:isoDensity})), therefore in this case there is no bifurcation. In
the generic case where induced catastrophes are present,
\eqref{eq:bifEquation} can be satisfied only if $\kappa(\theta)$
is an eigenfunction of $\mathbf{C}$. We know that the family of
functions $\cosb{2n\theta}$, $n \ge 1$, are eigenfunctions of
$\mathbf{C}$ with eigenvalues $\hat{c}_{2n}$, and therefore get a
set of bifurcation values for $\bar{N}$, one for each eigenvalue:
$N_{2n}^* = (-2 \hat{c}_{2n}+\hat{z}_0)^{-1}$. In addition, we
know that the isotropic solution must be stable as $G \to
-\infty$, because in this limit the microtubules have a vanishing
length and do not interact. Therefore, the relevant bifurcation
point is that for the lowest value of $G$, corresponding with the
most negative eigenvalue of $\mathbf{C}$ (see also section
\ref{sec:stability}). Assuming that the induced catastrophe
probability increases monotonically with the collision angle,
$\hat{c}_2$ is always the most negative eigenvalue, so
\begin{equation}
N^* = \frac{1}{-2 \hat{c}_2 +\hat{z}_0}.\label{eq:Nbifurcation}
\end{equation}

We now derive the location of this bifurcation point in terms of the control parameter $G$. Denoting
$\bar{N}=\bar{L}\bar{K}$, equation \eqref{eq:isoIntermediate1} can be transformed to $\bar{N}(1-\hat{z}_0 \bar{N})^2
= \bar{L}^3 $, into which we can substitute $G \bar{L}=(\hat{c}_0+\hat{z}_0)\bar{N}-1$ from equation
\eqref{eq:isoEqn1} and solve for $G$ giving
\begin{equation}
G^3 \bar{N}(1-\hat{z}_0 \bar{N})^2 = \left[(\hat{c}_0+\hat{z}_0)\bar{N}-1\right]^3 \label{eq:GNrelation}
\end{equation}
Combining this with the result \eqref{eq:Nbifurcation} yields
\begin{align}
G^* &= (-2\hat{c}_2)^{1/3}\left( \frac{\hat{c}_0}{-2\hat{c}_2} -1\right).\label{eq:bifG}
\end{align}
The implication is that the location of the bifurcation point as a function of the control parameter $G$ is
determined entirely by the eigenvalues of the induced catastrophe function $c(\theta)$. Like the density in the
isotropic phase, the location of the bifurcation point, this time perhaps more surprisingly, does not depend on the
presence or amount of zippering.

\subsection{Segment length and mesh size}

An attractive interpretation of the microtubule length density
$K(\theta)$ is that it represents the density of `obstacles' that
are pointing in the direction $\theta$ as seen by a microtubule
growing in the perpendicular direction. From the obstacle density
we can define a mesh size $\xi(\theta)$ - the average distance
between obstacles. Taking into account the geometrical factor
$\sinb{\theta}$, we obtain
\begin{equation}
\xi(\theta) = \left[ \frac{1}{\pi l_0 }
\integral{\left|\sinb{\theta -
\theta'}\right|K(\theta')}{\theta'}{0}{2\pi}\right]^{-1}.
\end{equation}
In the case of the isotropic solution, this simplifies to
$\bar{\xi} = \pi l_0/(4 \bar{K})$. Using this equality we can
derive an expression for the average microtubule length
$\bar{\Lambda}$ in the isotropic phase, expressed in units of the
mesh size. The length of each segment is given by $\bar{L}$ and
the number of segments per microtubule is given by $(1+\bar{Q})$,
so using \eqref{eq:isoIntermediate1} we find
\begin{equation}
\bar{\Lambda} = \frac{l_0 \bar{L}(1+\bar{Q})}{\bar{\xi}} = \frac{4
\bar{K}^{3/2}}{\pi}  \label{eq:defLambda}
\end{equation}
Inserting this result into \eqref{eq:isoDensity} provides the
relationship between $\bar{\Lambda}$ and $G$
\begin{equation}
G  = \left( \frac{4 }{\pi \bar{\Lambda}}
\right)^{\frac{1}{3}}\left(\frac{\pi \hat{c}_0 \bar{ \Lambda}}{4
}-1\right)\label{eq:relationGLambda}
\end{equation}
As was the case for the density, we see that the microtubule
length as a function of mesh size does not depend on the amount of
zippering. However, it should be noted that the mesh size is
defined through the average distance between single microtubules.
In real systems, zippering would naturally lead to bundling, which
in turn produces a system that has a larger mesh size between
bundles (see also the discussion).

Combining equations \eqref{eq:defLambda} and \eqref{eq:Nbifurcation}, the expression for $\bar{\Lambda}$ at the
bifurcation point becomes
\begin{align}
\bar{\Lambda}^* &= -\frac{2}{\pi \hat{c}_2} \label{eq:bifLambda}
\end{align}
Assuming a monotonically increasing induced catastrophe probability $P_{\textrm{c}}(\theta)$, we know that the
minimum value for $\hat{c}_2$ is reached when every collision at an angle larger than $45\degree$ leads to a
catastrophe. From \eqref{eq:bifLambda}, we see that this implies $\Lambda^* \ge 3/(2\sqrt{2})$, meaning that for a
bifurcation to occur, the microtubules need to be longer (sometimes much longer) than the mesh size, as is to be
expected.

Equation \eqref{eq:relationGLambda} can also provide an
interpretation of the length scale $l_0$. In the absence of
catastrophic collisions, we find
\begin{equation}
\bar{\Lambda}|_{\hat{c}_0=0} = \frac{4}{\pi}(-G^{-3}) =
\frac{4}{\pi}\left( \frac{l}{l_0} \right)^3,
\end{equation}
where $l=-1/g$ is the average length of the microtubules. $l_0$ is
therefore a measure of the microtubule length that is required to
enable a significant number of interactions ($\bar{\Lambda}=4/\pi$
for $l=l_0$). If the free microtubule length $l$ is (much) shorter
than $l_0$, the system is dominated by the (isotropic)
nucleations, keeping the system in an isotropic state. On the
other hand, when $l \gg l_0$, the interactions dominate and,
depending on the interaction functions, the system has the
potential to align.

\begin{figure*}[hbt]
  \begin{centering}
   \includegraphics{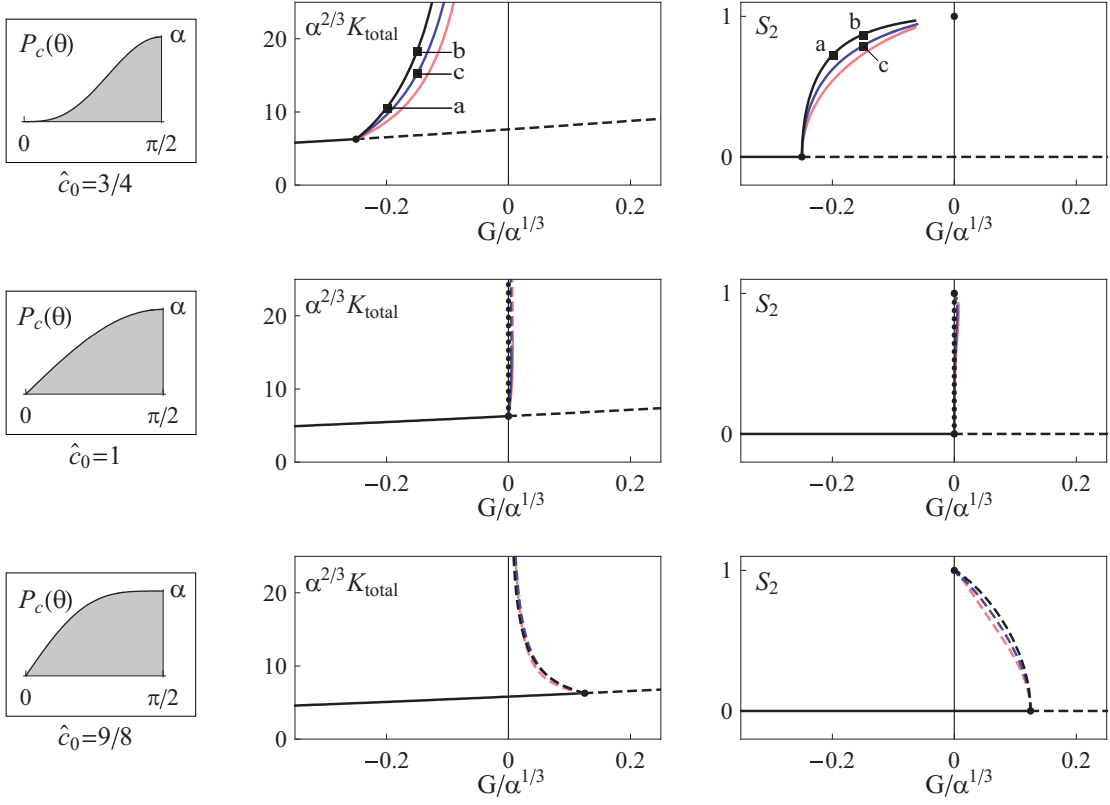}
   \caption{\label{fig:results-1}
Bifurcation diagrams for the simplified interaction functions using three different induced catastrophe parameters.
The figures on the left depict the probability $P_{\textrm{c}}(\theta)$ to induce a catastrophe upon collision,
along with the corresponding values of $\hat{c}_0$. The centre and right columns depict the corresponding
bifurcation diagrams as a function of $G$, expressed in terms of the total density $K_{\textrm{total}}$ and the 2D
nematic order parameter $S_2$, respectively, where $K_{\textrm{total}}=\int K(\theta) \mathrm{d}\theta$. The
isotropic solutions are by definition disordered, so $S_2=0$, and their density is computed from
\eqref{eq:isoDensity}. The bifurcation point is determined using \eqref{eq:bifG}, with $\hat{c}_2=-1/2$. For each
diagram, ordered solutions have been computed for $\hat{z}_0=0$ (black), $\hat{z}_0=1$ (blue) and $\hat{z}_0=10$
(red). The solutions have been computed using the method discussed in appendix \ref{sec:fourierprojection}. Solid
lines indicate stable solutions and dashed lines indicate unstable solutions (see also section \ref{sec:stability}).
Note that the case of $\hat{c}_0=1$ in the absence of zippering is a singular case where the stability cannot be
determined, because non-isotropic solutions only exist for $G=0$. This has been indicated by a dotted line. The
$S_2$-diagrams include the asymptotic limit point at $G=0$ with absolute ordering (at infinite density). The labels
\emph{a}, \emph{b} and \emph{c} indicate the parameter values of the results depicted in figure \ref{fig:results-2}.
The fact that the solutions for $S_2$ in the case $\hat{c}_0=\frac{3}{4}$ do not reach the asymptotic point
$(G=0,S_2=1)$ is a consequence of the slowdown in convergence of the path-following method as $G\uparrow 0$.}
 \end{centering}
\end{figure*}

\subsection{Ordered solutions for simplified interaction functions}\label{sec:minimalmodel}

To find solutions beyond the immediate vicinity of the bifurcation point, we are hampered by the fact that these
solutions are part of an infinite-dimensional solution space. In appendix \ref{sec:fourierprojection} it is shown
that the solutions can be constrained to a finite-dimensional space by restricting the interaction functions
$c(\theta)$ and $z(\theta)$ to a finite number of Fourier modes.

In this section, we will define a set of simplified interaction functions by restricting ourselves to Fourier modes
up to and including $\cos(4\theta)$. These modes provide us with just enough freedom for the model to exhibit rich
behaviour. Using the fact that $c(0)=z(0)=z(\pi/2)=0$, we find that $\hat{z}_2=0$ and that both $\hat{z}_4$ and
$\hat{c}_4$ are determined by the remaining parameters. Furthermore, we introduce an overall factor of $\alpha$ in
both equations, allowing us to set $\hat{c}_2=-1/2$, so that $c(\pi/2)=\alpha$. We thus obtain a system that is
fully specified by the parameters $\hat{c}_0$, $\hat{z}_0$ and $\alpha$.
\begin{subequations}\label{eq:minimalmodel}
\begin{align}
c(\theta)&= \alpha\left[ \frac{\hat{c}_0}{2} - \frac{1}{2} \cosb{2\theta} +\frac{1}{2}(1- \hat{c}_0)\cosb{4\theta} \right]\\
z(\theta)&=\alpha \left[\frac{\hat{z}_0}{2}(1- \cosb{4\theta})
\right]
\end{align}
\end{subequations}
For $\alpha=1$, $\hat{c}_0$ and $\hat{z}_0$ are the actual Fourier
coefficients of the interaction functions. Demanding that
$P_{\textrm{c}}(\theta)=c(\theta)/\sinb{\theta}$ is monotonically
increasing on the interval $[0,\pi/2]$ leads to the constraint
\begin{equation}
\frac{3}{4} \le \hat{c}_0 \le \frac{9}{8}
\end{equation}
and $\hat{z_0}$ is a positive real number. Of course, the total probability of zippering and catastrophe induction
may not exceed $1$, placing an upper bound on $\alpha$. In the absence of zippering, we have $\alpha \le 1$.

\begin{figure}[hbt]
  \begin{centering}
   \includegraphics{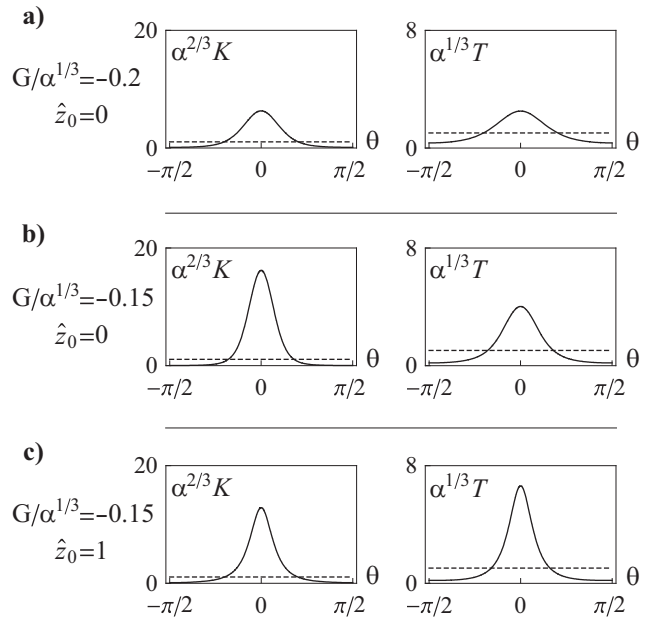}
   \caption{\label{fig:results-2}
Three stable ordered solutions that correspond to the points labelled (a), (b) and (c) in figure
\ref{fig:results-1}. The (unstable) isotropic solutions for the same parameter values are indicated with a dashed
line. The parameter values for (a) and (b) differ only in the value of $G$, whereas the parameter values for (b) and
(c) differ only in the value of $\hat{z}_0$. All results have been calculated using the method described in appendix
\ref{sec:fourierprojection}.}
 \end{centering}
\end{figure}

It should be noted that the value of $\alpha$ has no qualitative
effect on the results. This can be understood by realizing that
the set of equations \eqref{eq:setDimensionless} is invariant
under the substitutions
\begin{align}
\mathbf{C} &\to \alpha\mathbf{C}   & L &\to \alpha^{-1/3} L \nonumber\\
\mathbf{Z} &\to \alpha\mathbf{Z}   & K &\to \alpha^{-2/3} K \nonumber\\
G &\to \alpha^{1/3}G   & T &\to \alpha^{-1/3} T \nonumber
\end{align}
where the first substitution reflects the presence of $\alpha$ in
the definitions \eqref{eq:minimalmodel}. The existence of this
scaling relation implies that all functional dependencies between
any of these parameters and variables remain unchanged when
subjected to the inverse scaling. Explicitly, the relevant
parameters become $\mathbf{C}/\alpha$, $\mathbf{Z}/\alpha$ and
$\alpha^{-1/3}G$ and the variables $\alpha^{1/3}L$,
$\alpha^{2/3}K$ and $\alpha^{1/3}T$. With this in mind, we have
used the arbitrary choice $\alpha=1$ for our numerical
calculations, indicating the appropriate scaling on the axes of
figures \ref{fig:results-1} and \ref{fig:results-2}.

Equation \eqref{eq:bifG} indicates that, for the
simplified interaction functions, the bifurcation point
is located in the range
\begin{equation}
-\frac{1}{4} \le G^* \le \frac{1}{8}
\end{equation}
and from \eqref{eq:isoDensity} and \eqref{eq:bifLambda} we find
that $K^*=\alpha^{-2/3}$ and $\Lambda^* = 4/(\alpha \pi)$. We have
used the numerical procedure described in appendix
\ref{sec:fourierprojection} to determine the ordered solutions of
\eqref{eq:setDimensionless}, starting from the bifurcation point.
This has been done for nine different parameter values. For the
values of $\hat{c}_0$ we used the extreme values $3/4$ and $9/8$,
as well as $1$, the latter corresponding to $G^*=0$. For each of
these three cases, we have varied the zippering parameter
$\hat{z}_0$, choosing values of $0$, $1$ and $10$. Figure
\ref{fig:results-1} shows the results, depicting both the total
density of the system and the degree of ordering as a function of
$G$. The degree of ordering is measured by the order parameter
$S_2$, defined as
\begin{equation}
S_2[K(\theta)] = \frac{|\integral{\mathrm{e}^{i 2
\theta}K(\theta)}{\theta}{0}{2\pi}|}{\integral{K(\theta)}{\theta}{0}{2\pi}},
\end{equation}
which is the standard 2D nematic liquid crystal order parameter
(yielding 0 for a completely disordered system
and 1 for a fully oriented system).

\subsection{Stability of solutions}\label{sec:stability}

The bifurcation constraint \eqref{eq:bifEquation} indicates that
the space of bifurcating functions $\kappa_n$ is spanned by the
functions $\cosb{2n\theta}$ and $\sinb{2n\theta}$ for a given
value of $n \ge 1$. These solutions are therefore symmetric with
respect to an arbitrary axis that we choose to place at
$\theta=0$. Even after the restriction to this symmetry axis there
are still two solution branches emanating from the bifurcation
point, differing in the sign of the coefficient of the
perturbation. These branches correspond to solutions peaked around
$\theta=0$ and $\theta=\pi/(2n)$, respectively, that are identical
except for this rotation. The symmetry of these solutions
indicates that the bifurcation is of the pitchfork type. In figure
\ref{fig:results-2} we plot the solutions with a maximum at
$\theta=0$.

The presence of a pitchfork bifurcation implies a loss of
stability of the originating branch \cite{Golubitsky84}. In our
system, we know that the isotropic solution must be stable in the
limit $G\to -\infty$. Therefore, the local stability of the
isotropic solution is lost at the first bifurcation point (for the
lowest value of $G$), corresponding to the eigenfunction
$\cosb{2\theta}$. Because this eigenfunction is orthogonal to the
eigenfunctions related to the subsequent bifurcation points
($\cosb{2n\theta}, n > 2$), the stability of the unstable mode
will not be regained at any point along the isotropic solution and
the isotropic solution itself remains unstable for all higher
values of $G$. This also means that the solution branches
originating at further pitchfork bifurcations will be unstable
near the isotropic solution. In this paper we restrict ourselves
to the analysis of the first bifurcation point and the
corresponding ordered solution branch. Because the solutions on
this branch already have the lowest symmetry permitted by the
interaction functions, there are no further bifurcation points
along this branch.

Generically \citep[chap. IV]{Golubitsky84}, the branches of the
initial pitchfork bifurcation are stable for a supercritical
bifurcation (branches bending towards higher values of $G$) and
unstable for a subcritical bifurcation (branches bending towards
lower values of $G$). In addition, turning points in the
bifurcating branches generally correspond to an exchange of
stability \citep[page 22]{Iooss80}. This analysis allows us to
assign stability indicators to the bifurcation diagrams in figure
\ref{fig:results-1}, even in the absence of a detailed study of
the time-dependent equations \eqref{eq:MEall}.

\section{Discussion and conclusions}

Based on biological observations, we have constructed a model for the orientational alignment of cortical
microtubules. The model has a number of striking features. First of all, for a given set of induced catastrophe and
zippering probabilities ($P_{\textrm{c}}(\theta)$ and $P_{\textrm{z}}(\theta)$), it allows us to identify a single
dimensionless control parameter $G$, which is fully determined by the nucleation rate and intrinsic dynamics of
individual microtubules. This result by itself may turn out to be very useful in comparing different \emph{in vivo}
systems or the same system under different conditions or in different developmental stages. For increasing values of
$G$, the isotropic stationary solutions to the model show an increase both in density and in abundance of
interactions, as measured by the ratio of microtubule length to the mesh-size. Secondly, the bifurcation point, i.e.
the critical value of $G^{*}$ of the control parameter at which the system develops ordered stationary solutions
from the isotropic state, is determined solely by the probability of collisions between microtubules that lead to an
induced catastrophe.

Indeed, from the numerical solutions of the minimal model introduced in Section \ref{sec:minimalmodel} it appears,
perhaps surprisingly,  that the co-alignment of microtubules due to zippering events, if anything, diminishes the
degree of order. These results identify the ``weeding out'' of misaligned microtubules --- by marking them for early
removal by the induced switch to the shrinking state --- as the driving force for the ordering process. Finally, in
spite of not being able to directly assess the stability of the solutions in the time-domain, we have provided
arguments that stable ordered solutions are possible for the regime $G<0$, i.e. where the length of individual
microtubules is intrinsically bounded.

For $G>0$, individual microtubules have the tendency to grow
unbounded, unless they are kept in check by catastrophic
collisions. Although (locally) stable solutions may exist for
values of $G$ that are not too large, for every $G>0$ there exists
a class of aligned `runaway' solutions with diverging densities.
The computed ordered solutions, regardless of their stability,
converge to a point with $G=0$, for which the microtubules are
perfectly aligned ($S_2=1$) and the system is infinitely dense.
The existence of this point can be understood by the fact that the
alignment also serves to decrease the number of collisions, and in
the limit of a perfectly aligned system, the (relative) number of
collisions vanishes.

How realistic is the model presented? To answer this question we need to consider several known factors that have not been
included. First of all, microtubules typically can de-attach from their nucleation sites and then perform so called
treadmilling motion, whereby the minus-end shrinks at a more or less steady pace, which is small compared to both
the growth- and the shrinking speed of the more active plus end. In the case that no zippering occurs at all it is
relatively easy to show that the effect of treadmilling simply leads to a renormalization of the parameter $G$ and
the interaction functions $c(\theta)$ and $z(\theta)$, but leaving the qualitative behaviour of the model identical
to the one discussed here. When zippering does occur, one expects the treadmilling to enhance the degree of ordering in
an ordered state, as over time it ``eats-up'' the, by definition less ordered, initial segments of each
microtubule. This effect is also consistent with the observation in figure 5c that in the case with zippering the
active tips are on average more strongly aligned than the average segment. In fact, given that the comparison
between figures \ref{fig:results-2}b and \ref{fig:results-2}c also shows that, all else being equal, zippering
sharpens the orientational distribution of the active tips as compared to the case with no zippering, it is
conceivable that the combination of zippering and treadmilling could lead to more strongly ordered systems for the
same value of the control parameter.

Next it is known that in vivo severing proteins, such as katanin are active in, and crucial to, the formation of the
cortical array \cite{rollmecak2006}. Although in principle the effect of severing proteins could be included in the
model, it would present formidable problems in the analysis as well as introduce additional parameters into the
model for which precise data is lacking.

Another effect that has not been taken into account explicitly is microtubule bundling. Whenever a microtubule
zippers alongside another segment, they form a parallel bundle \cite{barton2008}. However, the coarse-grained nature
of our model precludes the formation of bundles and only allows for alignment of the segments. This means that a
microtubule that is growing in a different direction encounters each microtubule separately rather than as a single
bundle. It is to be expected that the catastrophe and zippering rates stemming from $N$ individual collisions will
be higher than those from a single collision with a bundle of $N$ microtubules. Hence, in realistic systems the
event rate is likely to be lower than that predicted by the model, or, in other words, corresponds to a lower value of the scaling
parameter $\alpha$. Bundles are also thought to be more than just adjacently aligned microtubules, because they may
be stabilized through association with bundling proteins that could potentially decrease the catastrophe rate of
individual microtubules within a bundle (see \cite{gaillard08}). This is a non-trivial effect that should be
considered separately and is likely to lead to an increased tendency to form an ordered structure.

Finally, our model implicitly assumes that there is an infinite supply of free tubulin dimers available for
incorporation into microtubules. Although there is no definite experimental evidence for this, it is reasonable to
assume that \emph{in vivo} there is a limit to the size of the free tubulin pool. Such a finite tubulin pool would
have marked consequences for the behaviour of the model, because the growth speed, and possibly also the nucleation
rate, are dependent on the amount of free tubulin, or equivalently the total density of microtubules $k_{tot}$. To a
first approximation the growth speed is given by $v^{+}(k_{tot}) = v^{+}(k_{tot}=0) \left(1-\frac{k_{tot}}{k_{max}}
\right)$ where $k_{max}$ the maximally attainable density when all tubulin is incorporated into microtubules. This
allows for stable states to develop even when $G(k_{tot}=0)>0$, because under this pool-size constraint the length
of individual microtubules will always remain bounded, and the system will settle into a steady state with
$G(k_{tot})<G(k_{tot}=0)$. This behaviour could provide a biologically motivated mechanism by which a solution with
a particular density is selected.

To see whether our model, in spite of its approximate nature,
makes sense in the light of the available data we first use the
collision event probabilities obtained by Dixit and Cyr
\cite{DixCyr04} (see figure \ref{fig:int}) to obtain an estimate
for the bifurcation value of the control parameter of $G^{*} =
-0.15$ for the case of Tobacco BY-2 cells. An ordered phase of
cortical microtubules should therefore be possible provided
$G>G^{*}$. Given the available data on the microtubule instability
parameters in this same system taken from Dhonukshe et al.\
\cite{Dhonukshe2003} and Vos et al.\ \cite{Vos+04} we would
predict using the definition (\ref{eq:Gdefinition}) that this
requires the nucleation rate of new microtubules to be larger than
0.05 min$^{-1}\mu$m$^{-2}$ (Dhonukse) and 0.01 min$^{-1}\mu$
m$^{-2}$ (Vos) respectively. Both these estimates for a lower
bound on the nucleation rate are reasonable as they imply the
nucleation of order $10^3$ microtubules in the whole cortex over
the course of the build-up towards full transverse order,
comparable to the number that is observed.

Finally, we should point out that our model so far only addresses the question of what causes cortical microtubules
to align with respect to each other. Given that in growing plant cells the cortical array is invariably oriented
transverse to the growth direction, the question of what determines the direction of the alignment axis with respect
to the cell axes is as, if not more, important from a biological perspective. We hope to address this
question, as well as the influence of some of the as yet neglected factors mentioned above, in future work.

\begin{acknowledgments}
The authors thank Kostya Shundyak, Jan Vos and Jelmer Lindeboom
for helpful discussions. SHT is grateful to Jonathan Sherratt for
his comments on the stability of solutions. RJH was supported by a
grant within the EU Network of Excellence ``Active Biomics''
(Contract: NMP4-CT-2004-516989). SHT was supported by a grant from
the NWO programme ``Computational Life Science''(Contract: CLS
635.100.003). This work is part of the research program of the
``Stichting voor Fundamenteel Onderzoek der Materie (FOM)'', which
is financially supported by the ``Nederlandse organisatie voor
Wetenschappelijk Onderzoek (NWO)''.
\end{acknowledgments}

\begin{appendix}
\section{Numerical evaluation of the ordered solutions}\label{sec:fourierprojection}

The solutions to the set of equations \eqref{eq:setDimensionless} lie in an infinite-dimensional solution space.
This creates significant hurdles for the numerical search for solutions. In this section, we will see that it is
possible to restrict the solutions to a finite-dimensional space by imposing constraints on the interaction
operators $\mathbf{C}$ and $\mathbf{Z}$. In addition, we present a method to follow the branch of ordered solution
in this finite-dimensional space, starting from the bifurcation point \eqref{eq:bifG}.

We start by reformulating the set of equations \eqref{eq:setDimensionless} by replacing $L(\theta)$ and $T(\theta)$
through the definitions
\begin{align}
S(\theta) &= \frac{1}{L(\theta)}, &
U(\theta)&=\frac{1}{K(\theta)}\left( \frac{T(\theta)}{L(\theta)}
-1 \right).
\end{align}
Following these substitutions, the interaction operators are all applied at the outermost level of the equations,
enabling us to make use of their properties in Fourier space. Explicitly, we obtain
\begin{align}
S(\theta)&=-G+\mathbf{C}\left[ K\right] \left( \theta \right)
+\mathbf{Z}\left[ K\right] \left( \theta
\right) \\
Q\left( \theta \right) &=\mathbf{Z}\left[ K\left( 1+Q\right)/S
\right] \left( \theta \right) \\
U(\theta) &= \mathbf{Z}\left[ (1+K U)/S  \right] (\theta)\\
\intertext{and} K(\theta) &= \frac{1+Q(\theta)}{S^2(\theta) - U(\theta)(1+Q(\theta))}.
\end{align}
Denoting the Fourier components of $S(\theta)$, $Q(\theta)$ and $U(\theta)$, by $\hat{s}_n$, $\hat{q}_n$ and
$\hat{u}_n$, respectively, the interacting microtubule equations reduce to a (potentially infinite) set of scalar
integral equations:
\begin{subequations}\label{eq:setNumerical}
\begin{align}
\hat{s}_{2n}&=-2 \delta_{n,0} G+ \frac{\hat{c}_{2n}+\hat{z}_{2n}}{\pi} \integral{\cosb{2 n \theta} K(\theta)}{\theta}{0}{2\pi} \\
\hat{q}_{2n} &= \frac{\hat{z}_{2n}}{\pi}\integral{\frac{\cosb{2 n
\theta} K(\theta)
\left( 1+Q(\theta)\right)}{S(\theta)}}{\theta}{0}{2\pi} \\
\hat{u}_{2n} &= \frac{\hat{z}_{2n}}{\pi} \integral{\frac{\cosb{2 n
\theta} (1+K(\theta) U(\theta))}{S(\theta)}}{\theta}{0}{2\pi}.
\end{align}
\end{subequations}

From the structure of these equations, we immediately see that we
can greatly reduce the dimensionality of the problem by setting a
number of Fourier coefficients $\hat{z}_{2n}$ and $\hat{c}_{2n}$
to zero. In other words, by restricting our space of interaction
functions $c(\theta)$ and $z(\theta)$, the problem can be reduced
to a finite number of scalar equations.

Applied to the simplified interaction functions introduced in section \ref{sec:minimalmodel}, we know that the sets
of stationary solutions form lines in the 8-dimensional phase space spanned by the variables $\{\hat{s}_0,
\hat{s}_2, \hat{s}_4, \hat{q}_0, \hat{q}_4, \hat{u}_0, \hat{u}_4\}$ and the parameter $G$. At least two of such
solution lines exist, one corresponding to the isotropic solution and the other to the ordered solution, and these
lines intersect at the bifurcation point.

Within this 8-dimensional space, we have used a numerical path-following method similar to the one described in
\cite{allgower2003,deuflhard1987} that follows the ordered solution branch by searching for a local minimum in the
root mean error of the constituent equations \eqref{eq:setNumerical}. The search for ordered solutions is initiated
at the bifurcation point, with coordinates
\begin{align}
S^* &=\frac{\hat{z}_0 -2 \hat{c}_2}{(-2\hat{c}_2)^{2/3}}, & Q^* &= - \frac{\hat{z}_0}{2 \hat{c}_2}, & U^* &=
\frac{\hat{z}_0}{(-2\hat{c}_2)^{1/3}},
\end{align}
so that in the case of our simplified interaction model
\begin{multline}
\{G,\hat{s}_0, \hat{s}_2, \hat{s}_4, \hat{q}_0, \hat{q}_4,
\hat{u}_0, \hat{u}_4\}_0 =\\
\{ c_0 -1,2 (\hat{z}_0 +1),0,0,2 \hat{z}_0,0,2  \hat{z}_0,0 \}.
\end{multline}
The initial instability affects only the $\cosb{2\theta}$ mode.
This mode only appears in the equation for $\hat{s}_2$ and the
remaining parameters are affected only by higher order
corrections. For this reason we choose the initial direction of
the path to be the unit vector in the $\hat{s}_2$-direction and
the path is traced from there.

\end{appendix}

\end{document}